\documentclass{article}
\usepackage{base/spconf,amsmath,graphicx}
\usepackage{bm,amssymb,amsthm,amsfonts,algorithmic,mathrsfs,multirow,enumitem,caption,xcolor,mwe,stfloats,booktabs,textcomp}
\usepackage{fancyhdr}
\usepackage{atbegshi,picture}

\makeatletter
\let\NAT@parse\undefined
\makeatother
\definecolor{citecolor}{HTML}{0071BC}
\definecolor{linkcolor}{HTML}{FF6347}
\usepackage[pagebackref=true, breaklinks=true, letterpaper=true, colorlinks, citecolor=citecolor, linkcolor=linkcolor, bookmarks=false]{hyperref}

\graphicspath{{pics/}}
\def\tX{\mathcal{X}}
\def\mX{\mathbf{X}}
\def\vx{\mathbf{x}}
\def\tmul{*_\mathscr{T}}
\def\t{\mathcal}

\definecolor{changecolor}{HTML}{38cb7d}
\newcommand{\yhao}[1]{\textcolor{black}{#1}}

\title{Dynamic MRI using Learned Transform-based Tensor Low-Rank Network (LT$^2$LR-Net)}
\name{Yinghao Zhang, Peng Li, Yue Hu\thanks{This work is supported by the National Natural Science Foundation of China under Grant 61871159 and Natural Science Foundation of Heilongjiang YQ2021F005.}}
\address{School of Electronics and Information Engineering, Harbin Institute of Technology, Harbin, China}
\begin{document}
%
\maketitle
%
\thispagestyle{fancy}
\rhead{}
\cfoot{\footnotesize This work has been accepted by the IEEE ISBI conference. Copyright may be transferred without notice, after which this version may no longer be accessible.}
\renewcommand{\headrulewidth}{0mm}
\begin{abstract}
While low-rank matrix prior has been exploited in dynamic MR image reconstruction and has obtained satisfying performance, tensor low-rank models have recently emerged as powerful alternative representations for three-dimensional dynamic MR datasets.
In this paper, we introduce a novel deep unrolling network for dynamic MRI, namely the learned transform-based tensor low-rank network (LT$^2$LR-Net). First, we generalize the tensor singular value decomposition (t-SVD) into an arbitrary unitary transform-based version and subsequently propose the novel transformed tensor nuclear norm (TTNN). Then, we design a novel TTNN-based iterative optimization algorithm based on the alternating direction method of multipliers (ADMM) to exploit the tensor low-rank prior in the transformed domain. The corresponding iterative steps are unrolled into the proposed LT$^2$LR-Net, where the convolutional neural network (CNN) is incorporated to adaptively learn the transformation from the dynamic MR dataset for more robust and accurate tensor low-rank representations.
Experimental results on the cardiac cine MR dataset demonstrate that the proposed framework can provide improved recovery results compared with the state-of-the-art methods.

\end{abstract}
\begin{keywords}
dynamic MRI, learned transform-based tensor low-rank, deep unrolling network
\end{keywords}
\section{Introduction}
\label{sec:intro}

Dynamic magnetic resonance imaging (MRI) plays an important role in multiple clinical applications including cardiac, perfusion, and vocal tract imaging. Dynamic MRI collects more information than static MRI, which is helpful in the detection of certain types of diseases (e.g., cardiovascular diseases). However, it is usually challenging to obtain dynamic MR images with high spatiotemporal resolution within clinically acceptable scan time. 

Low-rank Casorati matrix prior \cite{ref_ktslr} has been utilized to reconstruct dynamic MR images, but it \yhao{may} break the high-dimensional structure and thus motivates the low-rank tensor-based models \cite{ref_tensor, yaman2019low, he2016accelerated}. However these tensor-based models are usually based on the \yhao{Candecomp/Parafac (CP)} \cite{yaman2019low} or Tucker decomposition \cite{he2016accelerated, ref_wmnn2, ref_wmnn4} and \yhao{suffer} from the NP-hard issues\cite{friedland2018nuclear} and high-computational complexity\cite{ref_tensor}.

Compared with the CP and Tucker decomposition, the tensor singular value decomposition (t-SVD) \cite{ref_tsvd} has a simple decomposition form computed by the matrix SVDs of the frontal slices. Moreover, the tensor nuclear norm (TNN) \cite{ref_tnn1} induced in the t-SVD framework is the exact convex envelope of the corresponding tensor rank. 
The traditional t-SVD is based on the fast Fourier transform (FFT) along a specific dimension. Ai \cite{ref_tnn_dmri} and Zhang \cite{zhang2022TMNN} have applied the FFT-based t-SVD framework to enforce the tensor low-rank constraint for dynamic MR image reconstruction. However, the enhancement of performance is limited, which may be due to the inexact low-rank representation in the FFT domain. Wang et al. \cite{ref_ttnn2} proposed to utilize the low-rank properties under coupled transform of the spatial two-dimensional framelet transform and temporal FFT.
\yhao{Song et al. \cite{ref_ttnn} proposed to generalize the FFT into arbitrary unitary transformation along one specific dimension of the tensor. Subsequently, they introduced the transformed t-SVD and the transformed TNN.} 
\yhao{However, The unitary transformation proposed in \cite{ref_ttnn} is subject to one-dimensional, which may limit the capability to represent the low-rank structure of the data. In addition, it is quite challenging to obtain optimal performance using a predefined transformation.}

To address the abovementioned issues, we propose a novel learned transform-based tensor low-rank network (LT$^2$LR-Net) with application in dynamic MR reconstruction. 
In order to better characterize the low-rank representation in the transformed domain, we propose to utilize the convolutional neural network (CNN) to adaptively learn the transformation from the dynamic MR dataset.
Specifically, we first generalize the standard t-SVD into a transformed t-SVD framework based on \emph{arbitrary} unitary transform and propose a novel transformed tensor nuclear norm (TTNN). Then, we elaborately design a TTNN-based iterative algorithm for dynamic MR reconstruction via the alternating direction method of multipliers algorithm (ADMM). Finally, inspired by the deep unrolling strategy \cite{aggarwal2018modl,ref_DCCNN,ref_slrnet}, the TTNN-based iterative algorithm incorporated with CNN is unrolled into the supervised LT$^2$LR-Net, which enforces the low-rank constraint on the feature domain learned by the CNN.
Experimental results on a prospective Cine MR dataset (real-time OCMR \cite{ref_ocmr}) demonstrate the superior performance of the LT$^2$LR-Net over the state-of-art methods.

\section{TTNN: transformed tensor nuclear norm}
\label{TTNN}

In this paper, we denote tensors by Euler script letters, e.g., $\tX$, matrices by bold capital letters, e.g., $\mX$, vectors by bold lowercase letters, e.g., $\vx$, and scalars by lowercase letters, e.g., $x$. For a 3-way tensor $\tX \in \mathbb{C}^{n_1 \times n_2 \times n_3}$, we denote $\tX^{(i)}$ as the $i$th frontal slice $\tX(:,:,i),i=1,2,...,n_3$. 
Let $\hat{\tX}_\mathscr{T}$ be the tensor obtained via applying an \emph{arbitrary unitary} transform $\mathscr{T}$ on $\tX$, i.e., $\hat{\tX}_\mathscr{T} = \mathscr{T}(\tX)$ and $\tX = \mathscr{T}^H(\hat{\tX}_\mathscr{T})$.

The $\mathscr{T}$-product of $\mathcal{A} \in \mathbb{C}^{n_1 \times n_2 \times n_3}$ and $\mathcal{B} \in \mathbb{C}^{n_2 \times n_4 \times n_3}$ based on a unitary transform $\mathscr{T}$ is a tensor $\mathcal{C} \in \mathbb{C}^{n_1 \times n_4 \times n_3}$ \cite{ref_ttnn,ref_ttsvd}, which can be expressed as 
\begin{equation}
    \mathcal{C} = \mathcal{A} *_\mathscr{T} \mathcal{B} = \mathscr{T}^H(\hat{\mathcal{A}}_\mathscr{T} \times \hat{\mathcal{B}}_\mathscr{T}),
\end{equation}
where `$\times$' denotes the frontal slice-wise matrix product, i.e., $\left(\mathcal{A} \times \mathcal{B}\right)^{(i)}=\mathcal{A}^{(i)}\mathcal{B}^{(i)}$. Then, the transformed tensor singular value decomposition of $\tX \in \mathbb{C}^{n_1 \times n_2 \times n_3}$ can be derived as follows
\begin{equation}
    \tX = \mathcal{U} \tmul \mathcal{S} \tmul \mathcal{V}^H,
\end{equation}
where $\t{U}\in \mathbb{C}^{n_1 \times n_1 \times n_3}$ and $\t{V}\in \mathbb{C}^{n_2 \times n_2 \times n_3}$ are unitary tensors with respect to $\mathscr{T}$-product, and $\t{S}$ is a tubal diagonal tensor. Specifically, the $i$-th frontal slice of $\hat{\tX}_\mathscr{T}$ in the transformed domain can be expressed as 
\begin{equation}
    \hat{\tX}^{(i)}_\mathscr{T} = \hat{\t{U}}^{(i)}_\mathscr{T}\hat{\t{S}}^{(i)}_\mathscr{T}\hat{\t{V}}^{(i)H}_\mathscr{T},
\end{equation}
\yhao{which indicates that the transformed t-SVD is computed in the unitary $\mathscr{T}$ transformed domain via computing the matrix SVDs of every frontal slice of $\hat{\tX}_\mathscr{T}$.}
The transformed multirank of $\tX$ is a vector $\mathbf{r} \in \mathbb{R}^{n_3}$ with its $i$th entry being the rank of $\hat{\tX}^{(i)}_\mathscr{T}$. 
Following the derivation of the traditional TNN in \cite{ref_ttnn} and \cite{ref_tnn1}, 
we propose the novel TTNN of a tensor $\tX \in \mathbb{C}^{n_1 \times n_2 \times n_3}$, as the sum of the nuclear norms of all frontal slices of $\hat{\tX}_\mathscr{T}$ in the arbitrary unitary transformed domain, i.e., 
\begin{equation}
    \Vert \tX \Vert_{TTNN} = \sum_{i=1}^{n_3}\Vert \hat{\tX}^{(i)}_\mathscr{T} \Vert_*,
\end{equation}
which can also be considered as the convex envelope of the sum of the entries of the transformed multirank, similar to the original TNN \cite{ref_tnn1}. 

\section{The proposed LT$^2$LR-Net}
\label{LT$^2$LR-Net}
\subsection{The TTNN-based iterative algorithm}
\label{TTNN alg}

We denote the distortion-free dynamic MR image as $\mathcal{X} \in \mathbb{C}^{n_{x} \times n_{y} \times n_{t}}$, where $n_x$ and $n_y$ denote the spatial coordinates, and $n_t$ is the temporal coordinate. The data acquisition of dynamic MRI can be modeled as
\begin{equation}
  \mathbf b = A(\mathcal{X}) +\mathbf{n},
\end{equation}
where $\mathbf b \in \mathbb{C}^{m}$ is the observed undersampled $k$-space data, $A: \mathbb{C}^{n_{x} \times n_{y} \times n_{t}} \rightarrow \mathbb{C}^m$ is the Fourier undersampling operator, and $\mathbf{n} \in \mathbb{C}^{m}$ is the Gaussian distributed white noise.

The TTNN-based dynamic MR reconstruction model can be formulated as,
\begin{equation}
    \label{model}
    \min_{\tX} \frac 12 \Vert A(\tX)-\mathbf{b} \Vert_F^2+\lambda \Vert \tX \Vert_{TTNN},
\end{equation}
where $\lambda$ is the balancing parameter.
The optimization problem \eqref{model} can be converted into the following constrained problem by the variable splitting strategy, 
\begin{equation}
    \min_{\tX} \frac 12 \Vert A(\tX)-\mathbf{b} \Vert_F^2+\lambda \Vert \t{Z} \Vert_{TTNN}
    \ \ s.t. \  \t{Z} = \tX.
\end{equation}
The augmented Lagrangian function of the above problem is as follows:
\begin{equation}
    \begin{aligned}
    \mathcal{L}(\tX, \t{Z}, \t{W}) = \frac 12 \Vert A(\tX)-\mathbf{b} \Vert_F^2+\lambda \Vert \t{Z} \Vert_{TTNN} + \\
    <\t{W}, \t{X} - \t{Z}> + \frac {\mu}2 \Vert \t{Z} - \tX \Vert_F^2,
    \end{aligned}
\end{equation}
where $\t{W}$ is the Lagrangian multiplier. The above problem can be efficiently solved with the alternating direction method of multipliers algorithm (ADMM), which yields in solving the following subproblems:
\begin{small}
\begin{align}
    & \t{Z}_{n} = \min_{\t{Z}} \lambda \Vert \t{Z} \Vert_{TTNN} + \frac {\mu}2 \Vert \t{Z} - \tX_{n-1} - \t{L}_{n-1}\Vert_F^2, \label{z}\\
    & \tX_{n} = \min_{\tX} \frac 12 \Vert A(\tX)-\mathbf{b} \Vert_F^2 + \frac {\mu}2 \Vert \t{Z}_{n} - \tX - \t{L}_{n-1}\Vert_F^2, \label{x}\\
    & \t{L}_{n} = \t{L}_{n-1} - \eta (\t{Z}_{n}-\tX_{n}),
\end{align}
\end{small}%
\yhao{where $\t{L}=\t{W}/\mu$, and the subscript $n$ denotes the $n$-th iteration.} Finally, \eqref{model} can be efficiently solved using the following iterative steps:
\begin{small} 
\begin{equation}
    \label{iter alg}
    \begin{cases}
        \mathbf{Z}_{n}: &\t{Z}_{n} = \mathscr{D}_{{\lambda}/{\mu}, \mathscr{T}}(\tX_{n-1} + \t{L}_{n-1}) \\
        \mathbf{X}_{n}: &\tX_{n} = (A^HA + \mu)^{-1}(A^H(\mathbf{b})+\mu \t{Z}_{n} - \mu\t{L}_{n-1})\\
        \mathbf{L}_{n}: &\t{L}_{n} = \t{L}_{n-1} - \eta (\t{Z}_{n}-\tX_{n}),
    \end{cases}
\end{equation}
\end{small}%
where $\mathscr{D}_{{\lambda}/{\mu}, \mathscr{T}}$ denotes the \yhao{transformed tensor singular value thresholding} ($\mathscr{T}$-TSVT) operator, which can be derived according to the standard TSVT \cite{ref_tnn1}. \yhao{Furthermore, $\mathscr{T}$-TSVT can be factorized into $\mathscr{T}^H \rm{SVT}_{{\lambda}/{\mu}}\mathscr{T}$ where the operator $\rm{SVT}$ denotes the frontal slice-wise singular value thresholding w.r.t. ${\lambda}/{\mu}$ threshold. The subproblem \eqref{x} is a quadratic problem, which can be solved analytically.}

\subsection{The LT$^2$LR-Net framework}
\label{net framework}

\begin{figure}[t]
    \centering
    \includegraphics[scale=0.375]{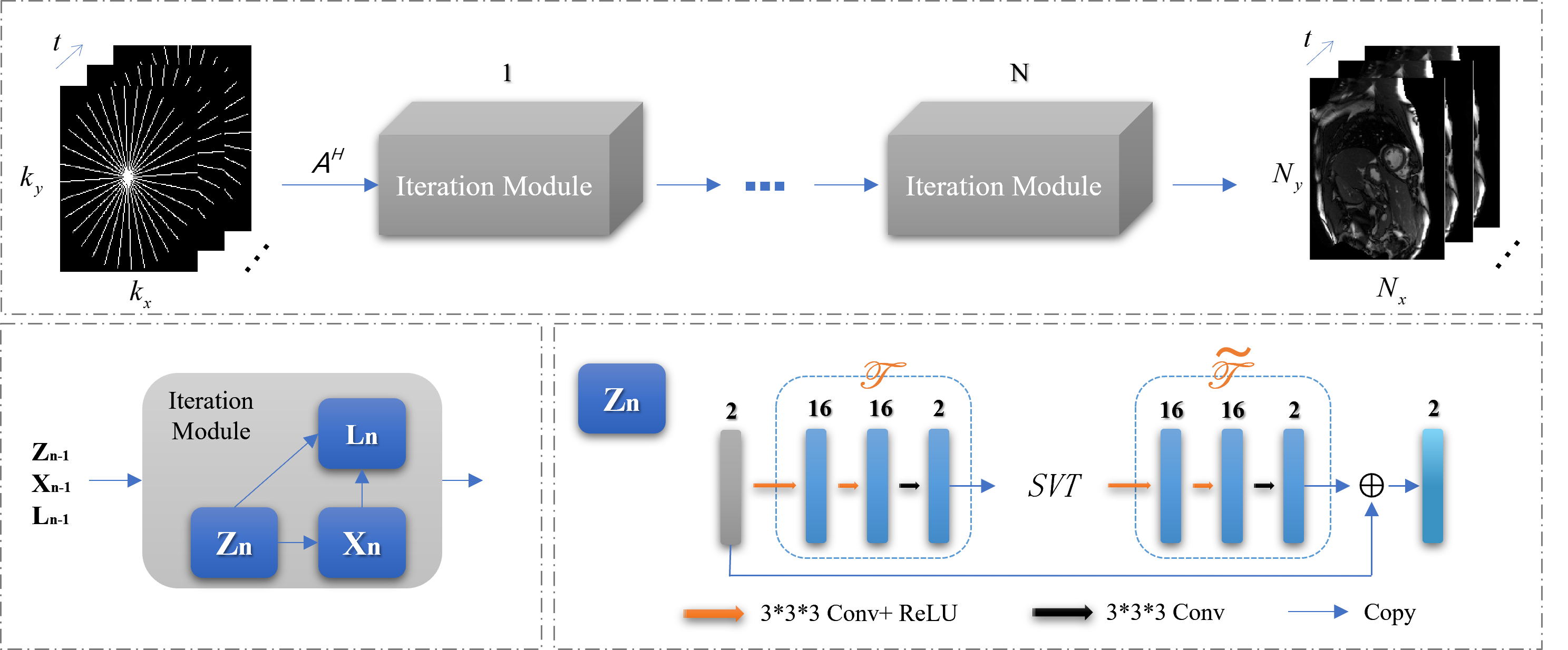}
    \caption{The LT$^2$LR-Net framework.}
    \label{fig_net}
    \vspace{-0.5cm}
\end{figure}

\yhao{In order to obtain the low-rank representation in the learned transformed domain, we unroll the TTNN-based iterative algorithm \eqref{iter alg} into the supervised LT$^2$LR-Net. The hyperparameters $\lambda, \mu, \eta$ are set as learned variables, and the TTNN-based transformation $\mathscr{T}$ are adaptively learned by CNN.}

The framework of LT$^2$LR-Net is shown in Fig.\ref{fig_net}. It unrolls \yhao{$N$} iterative steps of \eqref{iter alg} into $N$ iteration modules, each of which contains three blocks corresponding to the three subproblems, i.e., the transformed low-rank tensor prior block $\mathbf{Z}_n$, the reconstruction block $\mathbf{X}_n$, and the multiplier update block $\mathbf{L}_n$. To be exact, \yhao{the $\mathbf{X}_n$ and $\mathbf{L}_n$ blocks are consistent with \eqref{iter alg} except the hyperparameters $\lambda, \mu, \eta$ are learned. }For the $\mathbf{Z}_n$ block, LT$^2$LR-Net replaces $\mathscr{T}$ and $\mathscr{T}^H$ in $\mathscr{T}$-TSVT operation with two separate non-interacting CNNs, which have the same structure containing three 3*3*3 convolutional layers with 16, 16, and 2 channels. \yhao{The last convolutional layer of each CNN does not have ReLU to avoid truncating the negative part.}

Moreover, in order to additionally learn and exploit the implicit deep image priors, we relax the strict unitary constraint of these two CNN-learned transformations. The residual connection between the input and output has also been \yhao{incorporated \cite{ref_ISTANET}}. Finally, we straightly adopt the Mean Square Error (MSE) as the loss function of the proposed supervised network.

\yhao{The proposed LT$^2$LR-Net exploits the tensor low-rank prior in the feature domain learned by the CNN, which can be easily generalized to solve other problems, such as tensor completion \cite{ref_ttnn2} and robust PCA \cite{ref_ttnn} problems. }

\section{Experimental Results}
\label{results}

We evaluate the proposed LT$^2$LR-Net using the open-access real-time OCMR dataset \cite{ref_ocmr}. \yhao{We choose 51 slices of fully sampled data on the 3T Siemens MAGNETOM Prisma machine for training. 10 slices of the 3T data and the other 12 slices from the 1.5T Siemens Sola machine are selected for testing to evaluate the robustness and generalization ability of our LT2LR-Net. We crop the training data into 1099 images with the size of $144 \times 112 \times 16$ for data augmentation. }Note that all the multi-coil data from OCMR are combined into single-coil images, and the coil sensitivity maps are computed by ESPIRiT \cite{ref_multicoil}. 

For module configuration, we set LT$^2$LR-Net with 15 iteration modules, and we adopt the exponential decay learning rate with an initial learning rate of 0.001 and a decay of 0.95. The Adam optimization is adopted in the LT$^2$LR-Net training, and we use 50 epochs in the training of the proposed LT$^2$LR-Net and the unrolling methods under comparisons. All methods are implemented on Intel Xeon W-2123 CPU and Quadro GV100 GPU (32 GB memory).

In order to evaluate the performance of the proposed network, we compare the LT$^2$LR-Net with four iterative algorithms, namely, TNN \cite{ref_tnn1}, F2TNN \cite{ref_ttnn2}, k-t SLR \cite{ref_ktslr} and MNN3TV \cite{ref_wmnn2}. 
TNN and F2TNN are TNN-regularized methods using the traditional FFT and the predefined transform, respectively. The k-t SLR and MNN3TV utilize both the low-rank property and the total variation constraints, while the k-t SLR uses the low-rank matrix property and the MNN3TV is the low-Tucker-rank method.
Also, two state-of-art unrolling networks are considered in comparison, i.e., DC-CNN \cite{ref_DCCNN} and SLR-Net \cite{ref_slrnet}.

\begin{figure}[t]
    \centering
    \includegraphics[scale=0.25]{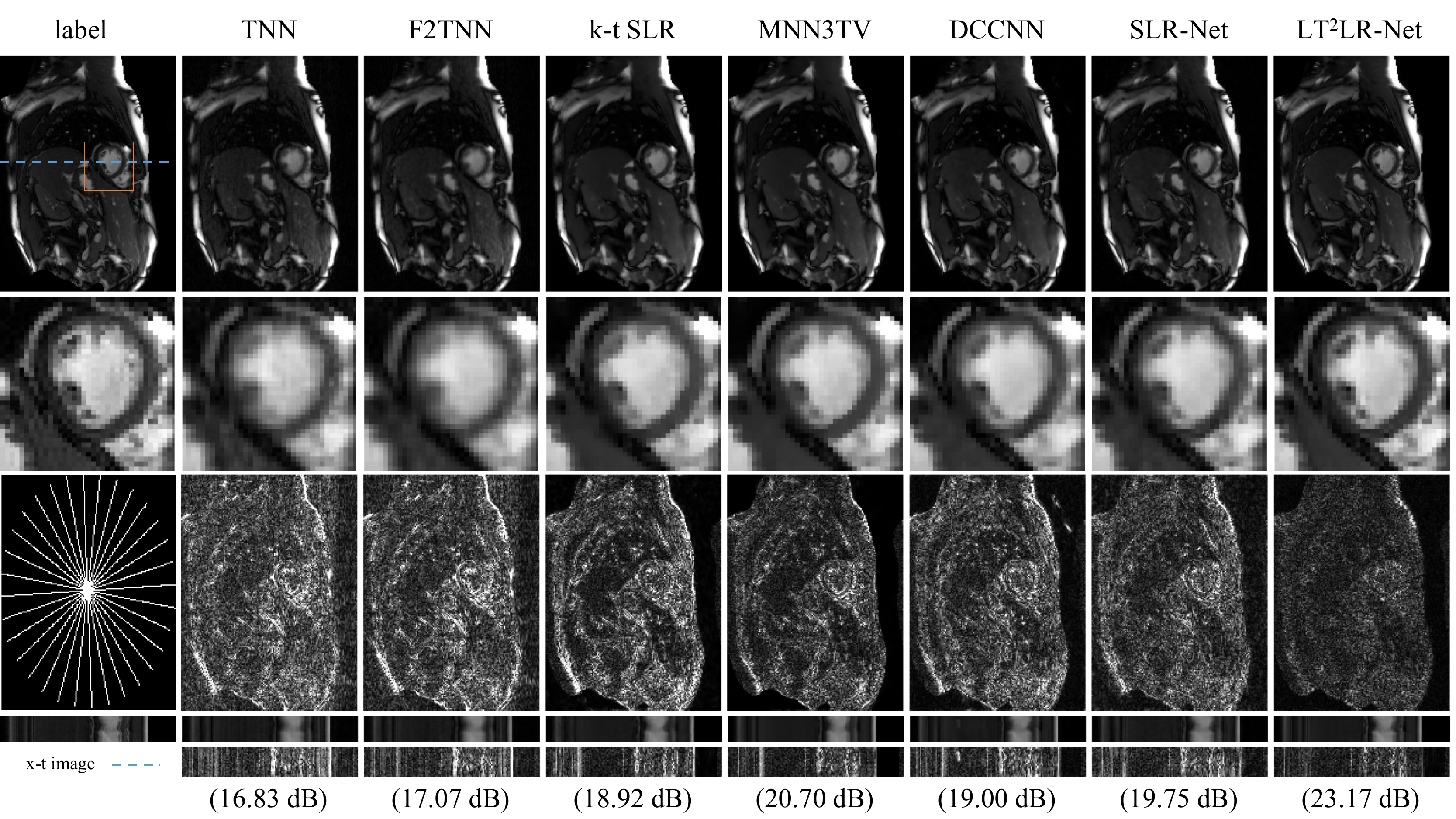}
    \caption{\small The reconstruction results of different methods under the pseudo-radial sampling pattern \cite{ref_ktslr} with 16 spokes. The first row shows the reconstruction images, and the second row shows the enlarged view of the heart regions marked by the orange box. The first image in the third row displays the sampling mask, while the others show the error maps. The fourth and fifth rows show the x-t images indicated by the blue dot line and their error maps. The reconstruction SNRs of different methods are listed in parentheses.}
    \label{fig_result}
    \vspace{-0.25 cm}
\end{figure}

The reconstruction results from measurements undersampled with pseudo-radial with 16 spokes are shown in Fig.\ref{fig_result}, where the acceleration factor (acc) is around 10.
It is observed that the reconstructed image by the proposed LT$^2$LR-Net retains the most image details and preserves the sharpest image edges compared with the other methods. In addition, the error maps show that the reconstruction performance of TLR-Net is significantly improved over the iterative methods and the unrolling networks. 

In addition, we also compare the proposed LT$^2$LR-Net with the other methods in terms of the SNR metric under the pseudo-radial and variable density sampling (vds) patterns, where 8, 16, 30 spokes are considered in radial pattern and the acceleration factors are set as 8, 10, 12 in vds pattern.
The quantitative evaluations are reported in Table.\ref{tab_result}. It is shown that LT$^2$LR-Net can provide the highest average SNR with the least parameters. Compared with the iterative algorithm, the proposed LT$^2$LR-Net provides substantially fast computation. The slightly longer time LT$^2$LR-Net takes compared with other unrolling networks may be due to the large-scale singular value decompositions. 
It is worth noting that the reconstruction results also indicate that our proposed LT$^2$LR-Net \yhao{outperforms} the basic TNN and the transformed tensor low-rank method, F2TNN, which illustrates the superiority of our learned transform-based tensor low-rank network.

\begin{table*}[htbp]
    \centering
    \caption{The average SNRs of the different methods under two sampling patterns (radial and vds) and six different sampling cases on the test dataset (Mean ± Standard deviation).}
    \scalebox{0.8}{
      \begin{tabular}{cccc|ccc|r|r}
      \toprule
      \multirow{2}[4]{*}{\textbf{methods}} & \multicolumn{3}{c|}{\textbf{radial / spokes}} & \multicolumn{3}{c|}{\textbf{vds / acc}} & \multicolumn{1}{c|}{\multirow{2}[4]{*}{\textbf{parameters}}} & \multicolumn{1}{c}{\multirow{2}[4]{*}{\textbf{time / s}}} \\
  \cmidrule{2-7}          & 8     & 16    & 30    & 8     & 10    & 12    &       &  \\
      \midrule
      TNN   & 12.39 ± 2.36 & 16.35 ± 2.68 & 19.66 ± 3.07 & 17.24 ± 2.32 & 15.39 ± 1.89 & 13.84 ± 1.74 & / & 20.5 \\
      F2TNN & 12.43 ± 2.25 & 16.57 ± 2.53 & 19.83 ± 2.92 & 17.66 ± 2.35 & 15.45 ± 1.90 & 13.89 ± 1.75 & / & 215.6 \\
      k-t SLR & 15.37 ± 2.14 & 18.54 ± 1.91 & 21.07 ± 2.31 & 18.83 ± 2.09 & 16.23 ± 1.95 & 14.65 ± 1.88 & / & 364.8 \\
      MNN3TV & 17.05 ± 2.28 & 19.93 ± 2.32 & 22.10 ± 2.50 & 19.94 ± 2.21 & 16.50 ± 1.94 & 14.91 ± 1.83 & / & 3472.8 \\
      DCCNN & 15.95 ± 1.79 & 18.53 ± 1.64 & 21.44 ± 2.25 & 20.37 ± 1.86 & 16.14 ± 1.80 & 14.48 ± 1.64 & 954340 & \textbf{1.3} \\
      SLR-Net & 16.14 ± 1.87 & 19.24 ± 1.76 & 21.27 ± 2.29 & 20.12 ± 1.73 & 16.05 ± 1.64 & 14.40 ± 1.52 & 293808 & 3.5 \\
      LT$^2$LR-Net & \textbf{19.12 ± 2.43} & \textbf{22.44 ± 2.60} & \textbf{24.08 ± 2.92} & \textbf{22.89 ± 2.63} & \textbf{16.89 ± 1.69} & \textbf{15.12 ± 1.66} & \textbf{259244} & 3.8 \\
      \bottomrule
      \end{tabular}}%
    \label{tab_result}%
  \end{table*}%

\vspace{-0.25 cm}
\section{Conclusion and discussion}
\label{conclude}

\yhao{We proposed a novel deep unrolling network that learns the transform-based tensor low-rank prior in dynamic MRI. Specifically, by generalizing the t-SVD decomposition to a unitary transformed version, we propose a novel TTNN and subsequently design a TTNN-based iterative algorithm based on ADMM to reconstruct the dynamic MR images. Then, the novel LT$^2$LR-Net is proposed via unrolling the iterative steps of the TTNN-based iterative algorithm. The proposed network adaptively exploits the tensor low-rank property in a CNN-learned transformed domain, which substantially outperforms the methods that use a fixed and predefined transformation. }Experimental results on OCMR dataset demonstrated the superior performance of the proposed LT$^2$LR-Net compared with the state-of-the-art methods.

\bibliographystyle{base/IEEEtran}
\footnotesize\bibliography{base/refs}

\end{document}